\documentclass[aps,prl,twocolumn,showpacs,superscriptaddress,groupedaddress]{revtex4} 

\usepackage{amsmath}
\usepackage{amssymb}
\usepackage{dcolumn}
\usepackage{bm}
\usepackage{braket}
\usepackage{float}
\usepackage{graphicx}
\usepackage{tikz}
\usepackage{slashed}
\usepackage{hyperref}
\begin{document}

\title{Entanglement Spectrum of a Random Partition:\\ Connection with the Localization Transition}

\author{Sagar Vijay}
\author{Liang Fu}
\affiliation{Department of Physics, Massachusetts Institute of Technology,
Cambridge, MA 02139, USA}

\begin{abstract} 
 We study the entanglement spectrum of a translationally-invariant lattice system under a random partition, 
 implemented by choosing each site  to be in one subsystem with probability $p\in[0, 1]$. 
We apply this random partitioning to a translationally-invariant (i.e., clean) topological state, and argue on general grounds that the corresponding entanglement spectrum captures the universal behavior about its \emph{disorder-driven} transition to a trivial localized phase. Specifically,  as a function of the partitioning probability $p$,  
the entanglement Hamiltonian $H_{A}$ must go through a topological phase transition driven by the percolation of a random network of edge-states.  As an example, 
we analytically derive the entanglement Hamiltonian for a one-dimensional topological superconductor under a random partition, and demonstrate that its phase diagram includes transitions between Griffiths phases. 
\end{abstract}

\maketitle

In recent years, systematic studies of quantum entanglement have greatly advanced our understanding of topological states of matter that cannot be adiabatically connected to a trivial product state. For example, topological entanglement entropy is directly related to the total quantum dimension of fractional quasi-particles \cite{levinwen, Kitaev_Preskill}. 
More recently, there has been a growing interest in utilizing the full entanglement spectrum to extract other universal properties \cite{Li_Haldane}, especially in chiral topological phases (e.g., quantum Hall states) or symmetry-protected topological phases (e.g., topological insulators). As a common feature, these phases have topologically-protected gapless excitations on a physical boundary. When the ground state is spatially cut into left and right halves, the low-lying part of the entanglement spectrum shares the same universal characteristics as the energy spectrum of these boundary excitations \cite{Li_Haldane, Edge_1, Edge_2, Edge_3, Edge_4, Edge_5, Edge_7}.      

A recent work studied the entanglement spectrum obtained from an extensive partition that divides a system into two {\it extensive} subsystems \cite{Hsieh_Fu}. For topological states that support gapless edge states, the corresponding entanglement spectrum was found to encode a wealth of information about the universal quantum critical behavior that would arise at its phase transition to a trivial direct-product state, despite the fact that the system under study itself is non-critical. It has been further shown that the entanglement spectra of extensive partitions can be directly computed from the matrix product state or tensor network representation of ground-state wavefunctions \cite{Hsieh_Fu_Qi}, which may offer insights into topological phase transitions \cite{Zhang, santos, hatsugai, Borchmann}.  

In this Letter, we study the entanglement spectum generated from a random partition that spatially bipartitions a system in a {\it probabilistic} manner: each physical site is chosen to be in subsystem $B$ (or $A$) with a probability $p\in[0,1]$ (or $1-p$). We apply this random partition to a translationally-invariant (i.e., clean) topological state, and argue on general grounds that the corresponding entanglement spectrum reproduces the universal behavior about its \emph{disorder-driven} transition to a trivial localized phase.  As an example, we analytically derive the form of the entanglement Hamiltonian for a clean one-dimensional topological superconductor under random partition, and establish the entanglement phase diagram as a function of probability $p$, finding agreement with the physical phase diagram of a disordered superconductor \cite{Huse_Motrunich_Damle}. 

We begin by considering a translationally-invariant topological state, which can be either a topological insulator/superconductor or a bosonic symmetry-protected topological phase.  
It has been shown \cite{Hsieh_Fu} that upon varying the geometry of $A$ and $B$ subsystems in an extensive partition,  the corresponding entanglement Hamiltonian undergoes a gap-closing transition that lies in the same universality class as the transition to a topologically trivial state realized by tuning the \emph{physical} Hamiltonian.  This intriguing connection is not a coincidence, but follows from the nature of topological phase transitions, which are driven by the percolation of gapless edge-states. For example, the transition from a quantum Hall insulator to a trivial insulator is described by the quantum percolation of chiral edge-states in a Chalker-Coddington network model \cite{Chalker_Coddington}. Extensive partitioning of a quantum Hall insulator precisely creates, in the low-lying part of the entanglement spectrum, a network of chiral edge states moving along the percolating borders between $A$ and $B$. This mapping explains why topological phase transitions and entanglement spectra of extensive partitions are intimately related. It further motivates us to study the random partitioning of a topological ground-state, for which the entanglement spectrum is expected to mimic the network model with randomness and thus connect with the localization transition.

We define the probabilistic partitioning of a clean, topological state $\ket{\Psi}$ as follows.  We independently choose each physical site in the full system to be in the $B$ subsystem with probability $p$; the remaining sites are defined to be in subsystem $A$. A partial trace of the density matrix over sites in subsystem $B$ yields a reduced density matrix for the $A$ subsystem
$\rho_A\equiv {\rm Tr}_B |\Psi \rangle \langle \Psi|$, which can be  interpreted as the \emph{thermal} density matrix at temperature $T = 1$ for an entanglement Hamiltonian $H_A$: $\rho_A \equiv e^{-H_{A}}$.  

Our goal is to determine the phase diagram of $H_{A}$ as a function of the partitioning probability $p$.  First, when $p \rightarrow 0$, the probabilistic partitioning yields a vanishingly small $B$ subsystem, with most sites belonging to the $A$ subsystem. In this limit, the ground-state of the entanglement Hamiltonian, denoted by $\ket{\psi_{A}}$, must share the same topological index as the original ground state $\ket{\Psi}$. As $p \rightarrow 1$, however, most sites become part of the $B$ subsystem, so that $\ket{\psi_{A}}$ becomes a trivial product state over the disjoint regions of the $A$ subsystem, and hence must be topologically trivial.  Since the topological character of $\ket{\psi_{A}}$  changes as we tune the partitioning probability $p$, we conclude that the entanglement Hamiltonian ${H}_{A}$ must go through a \emph{phase transition} at some critical partitioning probability $p = p_{c}$. Physically, the transition is driven by the percolation of a random network of gapless edge-states propagating around traced-out regions of the $B$ subsystem, as in the case of the aforementioned checkerboard-type extensive partition. Even though the original state $\ket{\Psi}$ is translationally-invariant, the probabilistic partitioning procedure introduces randomness into the entanglement Hamiltonian $H_A$, with the probability $p$ effectively tuning disorder strength.    

The phase diagram of the entanglement Hamiltonian as a function of $p$ satisfies additional constraints. For a given bipartition, the eigenvalue spectra of the reduced density matrices $\rho_A$ and $\rho_{B}$ are identical, though their Hilbert spaces are distinct. By definition,  for a fixed partitioning probability $p$,  
the $B$ subsystem is, on average, equivalent to the $A$ subsystem obtained with a partitioning probability $1-p$. Therefore, the ensemble-averaged spectra of the entanglement Hamiltonians ${H}_{A}(p)$ and ${H}_{A}(1-p)$ must be identical. As a result, the presence of a phase transition in the entanglement Hamiltonian with partition probability $p$ implies another transition at probability $1-p$. In the case where the topological index of $\ket{\Psi}$ cannot be evenly divided between two subsystems, as is the case for topological insulators with a $\mathbb{Z}_{2}$ index or quantum Hall insulators with an odd Chern number, we further expect that $H_A$ exhibits at least a topological phase transition at partitioning probability $p = 1/2$, when the two subsystems are equivalent on average.  

\begin{figure}
\includegraphics[trim = 0 0 0 0, clip = true, width=0.45\textwidth, angle = 0.]{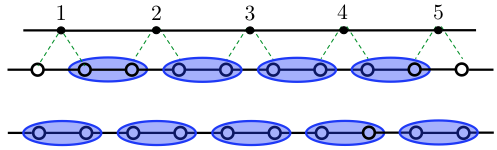}
\caption{The two ground-states of the Kitaev $p$-wave superconductor with dimerization of Majorana fermions \emph{across} (top) and \emph{within} (bottom) lattice sites, corresponding to a TSC and a trivial $p$-wave superconductor, respectively.}\label{fig:Kitaev_Chain}
\end{figure}

For the remainder of the paper, we apply our random partitioning procedure to Kitaev's model \cite{Kitaev} for
a clean one-dimensional $p$-wave superconductor, extract the phase diagram of the entanglement Hamiltonian as a function of partitioning probability, and demonstrate its correspondence with a disordered superconductor.  
The Kitaev model is described by the Hamiltonian
\begin{align}
H = &-w\sum_{n}\left(c^{\dagger}_{n+1}c_{n} + \text{h.c.}\right) + \sum_{n}\left(\Delta c^{\dagger}_{n+1}c^{\dagger}_{n} + \text{h.c.}\right)\nonumber\\
& - \mu\sum_{n}\left(c^{\dagger}_{n}c_{n} - \frac{1}{2}\right)
\end{align}
with fermion operators $c_{n}$, $c^{\dagger}_{n}$ satisfying canonical anti-commutation relations.  
To simplify the calculation below, we restrict ourselves to the case where $w = \Delta \in \mathbb{R}$.  Introducing two species of Majorana fermions at each lattice site $\gamma_{n} \equiv c_{n} + c^{\dagger}_{n}$ and $\chi_{n} \equiv (c^{\dagger}_{n} - c_{n})/i$, we may write the Hamiltonian as 
\begin{align}\label{eq:K_Hamiltonian}
{H} = \frac{iw}{2}\sum_{n}\left[\eta\gamma_{n}\chi_{n} + \chi_{n}\gamma_{n+1}\right]
\end{align}
where the dimensionless parameter $\eta \equiv \mu/2w$ distinguishes the topologically trivial strong pairing phase ($|\eta| > 1$) and topologically 
non-trivial weak pairing phase ($|\eta| < 1$). 
As shown in Figure \ref{fig:Kitaev_Chain}, in the topological superconductor (TSC) phase,  Majorana fermions couple more strongly {across} two adjacent lattice sites than within a lattice site, leading to two unpaired Majorana fermions at the ends of the chain. In the extreme limit when $\eta = 0$ in the topological phase, pairs of Majorana fermions within lattice sites completely decouple. We will refer to this special point $\eta=0$ as the \emph{Kitaev limit} of the Hamiltonian (\ref{eq:K_Hamiltonian}), which proves to be a useful starting point for our analysis below. 

Let $\ket{\Psi(\eta)}$ be the ground-state of the Kitaev Hamiltonian with parameter $|\eta|<1$ in the topological regime.  Applying the random partitioning procedure to this ground-state, we trace over  \emph{physical} lattice sites (each of which contains two Majorana fermions) with probability $p$, and obtain an entanglement Hamiltonian $H_{A}(\eta; p)$.  
Clearly, $H_A(\eta; p)$ contains random couplings between sites in the $A$ subsystem, and mimics the physical Hamiltonian of a disordered superconductor, with the partitioning probability $p$ playing the role of disorder strength.

To derive $H_A(\eta; p)$, we note that the entanglement Hamiltonian of an $N$-site free fermion system such as the Kitaev model can only contain fermion bilinear terms \cite{Peschel}. 
The spectrum of the entanglement Hamiltonian can be determined from the correlation matrices in the original ground-state $C_{nm} \equiv \braket{\Psi\,|\,c^{\dagger}_{n}c_{n}\,|\,\Psi}$ and $F_{nm} \equiv \braket{\Psi\,|\,c^{\dagger}_{n}c^{\dagger}_{m}\,|\,\Psi}$, by solving the eigenvalue problem \cite{Peschel}:
\begin{align}
(2\hat{C} - 2\hat{F} - 1)(2\hat{C} + 2\hat{F} - 1)\phi_{\ell} = \tanh^{2}\left(\frac{\epsilon_{\ell}}{2}\right)\phi_{\ell}
\end{align}
where $\epsilon_{\ell}$ is an eigenvalue of the entanglement Hamiltonian, with eigenvector $\phi_{\ell}$.  Rewriting the complex fermions in terms of Majorana operators, we may define the $2N \times 2N$ skew-symmetric correlation matrix $\hat{\Gamma}$ for the Majorana fermions which has eigenvalues $\pm\tanh(\epsilon_{\ell}/2)$ \cite{Peschel_2, Supp_Material}.  Since we are interested in the low-lying part of the entanglement spectrum $\epsilon_{\ell}\rightarrow 0$, $\tanh(\epsilon_{\ell}/2) \rightarrow \epsilon_{\ell}/2$ and hence $\hat{\Gamma}$ satisfies:
\begin{align}
2\hat{\Gamma}\phi_{\ell} \approx \pm\epsilon_{\ell}\phi_{\ell} 
\end{align}
Therefore, the correlation matrix for Majorana fermions in the original ground-state is {\it equivalent} to the entanglement Hamiltonian acting on low-lying states $\epsilon_{\ell} \rightarrow 0$ in the entanglement spectrum.  By building the correlation matrix for Majorana fermions in the ground-state $\ket{\Psi(\eta)}$, we may now construct the entanglement Hamiltonian for the Kitaev model after a random partition.

For arbitrary $\eta$, performing a random partition will generally produce an entanglement Hamiltonian with highly non-local couplings, due to the non-vanishing correlations between distant Majorana fermions in the ground state. However, for a sufficiently small $|\eta|$, i.e., when the system is close to the Kitaev limit, we may derive the form of $H_{A}(\eta; p)$ \emph{analytically}. 
Let us first consider the case $\eta=0$, when pairs of Majorana fermions decouple. A single cut between two adjacent lattice sites then produces, in the entanglement spectrum of the $A$ subsystem, an unpaired Majorana fermion at the end of $A$. Aside from this, the entanglement spectrum at $\eta=0$ is identical to the energy spectrum (properly normalized) of decoupled Majorana pairs in the Kitaev Hamiltonian. Therefore, performing a random partition with several cuts will yield an $A$ subsystem that consists of disjoint segments, each of which hosts unpaired Majorana fermions at the two ends. 
 
We now explicitly construct the entanglement Hamiltonian $H_{A}(\eta; p)$ near the Kitaev limit  by perturbing away from $\eta=0$. As one may expect, a small $\eta$ induces a small coupling between the unpaired Majorana fermions at ends of disjoint segments with the rest of the $A$ subsystem. By an analytical calculation \cite{Supp_Material}, we find that couplings between two Majorana fermions in $H_{A}(\eta; p)$ decrease exponentially with their separation in the original lattice. Therefore, it suffices  to include \emph{nearest-neighbor} couplings within subsystem $A$ only in 
$H_{A}(\eta; p)$.  

\begin{figure}
\includegraphics[trim = 14 5 155 0, clip = true, width=0.49\textwidth, angle = 0.]{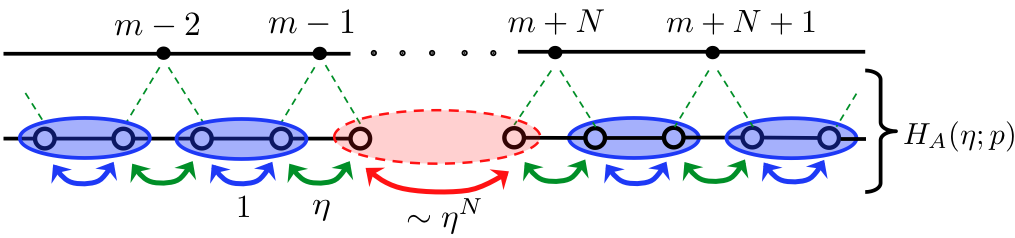}
\caption{The entanglement Hamiltonian $H_{A}(\eta; p)$, derived in the Kitaev limit $|\eta| << 1$, consists of several couplings between adjacent Majorana fermions in the subsystem $A$.  Blue and green hoppings between nearest-neighbor Majorana fermions appear with dimensionless coupling $1$ and $\eta$, respectively.  Tracing over $N$ lattice sites induces a coupling $O(\eta^{N})$ between the dangling Majorana modes on the adjacent chains in the $A$ subsystem. }\label{fig:Entang_Hamiltonian}
\end{figure}

Two types of nearest-neighbor couplings appear in $H_{A}(\eta; p)$. 
First, to leading order in $\eta$, couplings belonging to a connected sequence of sites in the $A$ subsystem are identical to those appearing in the original Kitaev Hamiltonian, after a proper normalization.  
Second, couplings between Majorana fermions belonging to different segments in the $A$ subsystem are computed from their two-point correlation function. If a series of $N$ consecutive lattice sites $-$ between sites $m - 1$ and $m + N$ $-$ are determined to be within the $B$ subsystem and traced over in the random partition, a coupling will be induced between the Majorana fermions at the right and left edges of the two lattice sites, which is found to be proportional to $C(N) \equiv \langle \Psi(\eta)\,|\,i\chi_{m-1}\gamma_{m+N}\,|\,\Psi(\eta)\rangle$.  
An explicit calculation \cite{Supp_Material} yields the result that at long distances, i.e. large $N$, 
\begin{align}
C(N) \sim \eta^{N}/\sqrt{N} + O(\eta^{N+1})
\end{align}
We then conclude that the entanglement Hamiltonian takes the form:
\begin{align}
{H}_{A}(\eta; p) = \frac{iw}{2}\sum_{n\in A}\eta\gamma_{n}\chi_{n} + \frac{iw}{2}\sum_{m > n \in A} f_{nm}\chi_{n}\gamma_{m} \label{ha}
\end{align}
where $f_{nm}$ has non-zero elements $f_{n, n+1} = 1$ and $f_{nm} = C(m - n - 1)$ if $n$ and $m$ label lattice sites at the right and left edges of two adjacent segments in the $A$ subsystem. 
The couplings in the entanglement Hamiltonian are illustrated in Fig. \ref{fig:Entang_Hamiltonian}.

We now demonstrate that the topological character of the entanglement Hamiltonian (\ref{ha}) changes at a critical partitioning probability $p = 1/2$.  Specifically, we demonstrate that when $p < 1/2$ the entanglement Hamiltonian in the $A$ subsystem supports an exponentially localized edge-state $\ket{\psi}$ at zero energy and corresponds to a topological superconductor phase, but that such a state does not exist for $p > 1/2$. 

To see this, we first note that a random partitioning of a chain with probability $p$ will produce an $A$ subsystem that consists of clusters of lattice sites. Let us now introduce a boundary in the $A$ subsystem and explicitly construct a 
zero-energy state of the entanglement spectrum under a random partition. Recall that an exact zero-energy left-boundary eigenstate of the translationally-invariant Kitaev Hamiltonian (\ref{eq:K_Hamiltonian}) takes the form $\ket{\psi} = (1, 0, \eta, 0, \eta^{2}, 0, \ldots)$, with $|\eta| < 1$, in the basis of Majorana sites on a semi-infinite chain \cite{Kitaev}.  We find a similar zero-energy state can be obtained for the entanglement Hamiltonian (\ref{ha}) of the $A$ subsystem, which consists of consecutive clusters of sites of lengths $\{\ell_{k}\}$, each separated by distance $\{d_{k}\}$. An edge-state of the entanglement Hamiltonian now takes the form:
\begin{align}\label{eq:Entang_Ground_State}
\ket{\psi_{A}} \propto &\Bigg(\underbrace{1, 0, \eta, 0, \ldots, \eta^{\ell_{1}}, 0,}_{\text{1st cluster of sites}}
\,{\frac{\eta^{\ell_{1}}}{C(d_{1})}, 0, \ldots, \frac{\eta^{\ell_{1} + \ell_{2}}}{C(d_{1})}, 0,\ldots}
\nonumber\\
&\underbrace{\prod_{k = 1}^{N-1}\frac{\eta^{\ell_{k}}}{C(d_{k})}, 0, \ldots, \eta^{\ell_{N}}\prod_{k = 1}^{N-1}\frac{\eta^{\ell_{k}}}{C(d_{k})}, 0}_{N^{\mathrm{th}} \text{ cluster of sites}}, \ldots \Bigg)
\end{align}
in a basis of Majorana sites in $A$.  From the form of the two-point function computed previously we see that amplitude for the edge-state on the first Majorana site in the $N$th cluster of the $A$ subsystem is given by:
\begin{align}
\psi_{N} = \prod_{k = 1}^{N-1}\frac{\eta^{\ell_{k}}}{C(d_{k})} \sim \prod_{k = 1}^{N-1}\eta^{\ell_{k} - d_{k}} = \eta^{\,\sum_{k}\ell_{k} - \sum_{k}d_{k}}
\end{align}   
Now, if we consider the amplitude at the end of the chain, we see that $\psi_{N} \rightarrow \eta^{L_{A} - L_{B}}$ where $L_{A}$ and $L_{B}$ are the sizes of the $A$ and $B$ subsystems, respectively. For a system of size $L$, regardless of the probability distributions for the lengths $\{\ell_{k}\}$ and $\{d_{k}\}$, the sizes of the two subsystems are determined from the partitioning probability to be $L_{A} = (1-p)L$ and $L_{B} = pL$ in the thermodynamic limit $L\rightarrow +\infty$ so that
\begin{align}
\psi_{N} \rightarrow \left[\eta^{1 - 2p}\right]^{L}
\end{align}
When $p < 1/2$ we observe that the state (\ref{eq:Entang_Ground_State}) is an exact zero-energy eigenstate of the entanglement Hamiltonian. When $p > 1/2$, however, the amplitude at the end of the chain diverges and the above state becomes non-normalizable for an infinite set of clusters.

The above calculation of an edge-state immediately implies that the entanglement Hamiltonian $H_A$ changes from being topologically non-trivial at partitioning probability  $p<1/2$ to trivial when $p >1/2$, 
and hence must be critical at the point $p=1/2$. This transition can also be understood by integrating out the Majorana fermions in the interior of the clusters in the $A$ subsystem and constructing an \emph{effective} entanglement Hamiltonian $H_{A}^{\mathrm{eff}}$ acting exclusively on the dangling Majorana modes at the ends of each cluster. In this case, $H_{A}^{\mathrm{eff}}$ will describe a dimerized Majorana fermion chain, in which two adjacent Majorana fermions correspond to sites separated by lengths $\{ ... \ell_k , d_{k}, \ell_{k+1}, d_{k+1}... \}$  in the original lattice. The nearest-neighbor hopping in $A$, which is 
proportional to the corresponding correlation function in the ground-state, is determined by the lengths $\{ d_{k}\}$ for intra-cluster hoppings or $\{ \ell_{k+1} \}$ for inter-cluster hoppings.    
At $p=1/2$, the $A$ and $B$ subsystems are equivalent on average, so that the length distributions $\{\ell_k\}$ and $\{d_k\}$ are identical, and   
the ensemble of $H_{A}^{\mathrm{eff}}$ is translationally-invariant, instead of dimerized. The corresponding ground state of a one-dimensional Majorana fermion chain is well-known to be critical \cite{Fulga}.  

We now demonstrate that in the vicinity of $p = 1/2$,  the entanglement Hamiltonian is in Griffiths phases, characterized by a singularity in the density of states at zero energy due to the proliferation of  segments of the topologically-ordered or trivial phase, respectively. Recall that when $p < 1/2$, near $p = 1/2$, the characteristic size of clusters in the $A$ subsystem is larger than that of the $B$ subsystem.  Then, the dangling Majorana modes on adjacent clusters in the $A$ subsystem, separated by distance $x$ will mix to form localized bound-states with finite energy $\epsilon \sim \exp[c\,x \ln |\eta|]$, with $c > 0$ a constant.  Since the probability of such a configuration of sites in the $A$ subsystem is $ p^{x}(1-p)^{2}$, the contribution of these low-energy modes to the density of states in the entanglement ground-state is \cite{Huse_Motrunich_Damle}:
\begin{align}\label{eq:DOS}
\rho(\epsilon) = \int_{0}^{\infty} dx \,  (1-p)^{2}p^{x} \,\delta(\epsilon - e^{c\,x \ln|\eta|}) \propto \frac{1}{\epsilon^{1-\beta(p)}}
\end{align}
with the non-universal exponent $\beta(p) \equiv \ln(p)/c\,\ln|\eta|$. The power-law singularity in the density of states signals the presence of a Griffiths region for an entanglement ground-state with $p$ near $1/2$, due to the proliferation of low-energy configurations of Majorana edge-modes dimerizing across lattice sites.  $p > 1/2$ also correspond to a Griffiths phase, with exponent $\beta(1-p)$  due to Majorana modes at the ends of the \emph{same} chain forming bound-states with exponentially small energy. The two Griffiths phases at $p<1/2$ and $p>1/2$ are both characterized by a power-law singularity in the density of states at zero energy, but are topologically distinct, as shown by the presence and absence of zero-energy Majorana fermion at the boundary.  

To summarize, we have introduced a random partitioning scheme to study the disorder-driven quantum critical behavior of a topological phase; applying this procedure to the one-dimensional $p$-wave superconductor yields an interesting phase diagram, consisting of two topologically distinct Griffiths phases separated by a critical point.  The random-partitioning scheme may be naturally used to study spin-chains \cite{Fisher, Yang}, as well as higher-dimensional systems to numerically extract critical exponents of disorder-driven phase transitions, such as localization transitions in all symmetry classes of non-interacting topological phases \cite{Classification_1, Classification_2}.  It might also be interesting to study the entanglement spectrum of fractional topological phases under a random partition, although the connection with topological phase transitions appears to be less direct.  

\acknowledgments
\emph{Acknowledgment:} We thank Tim Hsieh for helpful discussions. This work is supported by DOE Office of Basic Energy Sciences, Division of Materials Sciences and Engineering under award DE-SC0010526.

\section{Supplementary Material}

In this section,  we briefly review the calculation of the entanglement Hamiltonian for the ground-state of a free-fermion system as presented in \cite{Peschel_SM, Peschel_2_SM} and compute the two-point correlation function between Majorana fermions on even and odd Majorana sites in the ground-state of the Kitaev chain \cite{Kitaev_SM}, in the limit that the coupling $\eta \equiv \mu/2w$ is small.  In the main text, this calculation is used to construct the associated entanglement Hamiltonian, which is directly proportional to the correlation matrix for the Majorana fermions in the ground-state, when acting on the low-lying states in the entanglement spectrum.

\appendix
\section{The Entanglement Hamiltonian from a Bipartition of a Free-Fermion Ground State}
For a free-fermion Hamiltonian, the entanglement Hamiltonian obtained after a bipartition of the ground-state $\ket{\Psi}$ may be determined entirely from the two-point correlation functions in the ground-state.   For a free-fermion system, any $n$-point correlation function can be written as a sum over two-point functions by WickÕs Theorem.  Furthermore, the reduced density matrix $\rho_{A}$ is defined to reproduce all observables in the $A$ subsystem.  Therefore, for a fermion number-conserving, free-fermion Hamiltonian, the reduced density matrix must take the form:
\begin{align}
\rho_{A} = \frac{e^{-\mathcal{H}_{A}}}{Z} \hspace{.3in} \mathcal{H}_{A} \equiv \sum_{i,\, j \in A} h_{ij} c^{\dagger}_{i}c_{j}
\end{align}
with $Z \equiv \mathrm{Tr}\left[e^{-\mathcal{H}_{A}}\right]$ and fermion operators $c_{i}$, $c_{j}^{\dagger}$ satisfying canonical anti-commutation relations.  In this case, we may explicitly diagonalize the entanglement Hamiltonian $\mathcal{H}_{A}$, and express the correlation matrix $C_{nm} \equiv \braket{c^{\dagger}_{n}c_{m}}$ (with $n, m \in A)$ evaluated in the ground-state in terms of the eigenvalues $\{\epsilon_{j}\}$ and eigenvectors $\{\phi_{j}\}$ of $\mathcal{H}_{A}$ as \cite{Peschel_SM}:
\begin{align}
C_{nm} = \mathrm{Tr}\left[\rho_{A}c^{\dagger}_{n}c_{m}\right] = \sum_{j}\phi^{*}_{j}(n)\phi_{j}(m) \left(e^{\epsilon_{j}} + 1\right)^{-1}
\end{align}
In matrix form, this is equivalent to the statement that the entanglement Hamiltonian may be written as $(\mathcal{H}_{E})^{T} = \ln[(1-\hat{C})/\hat{C}]$, and that $\hat{C}$ solves the eigenvalue problem \cite{Peschel_2_SM}:
\begin{align}
(1 - 2\hat{C})\phi_{j} = \tanh\left(\frac{\epsilon_{j}}{2}\right)\phi_{j}
\end{align}

A similar calculation may be performed for a free-fermion Hamiltonian defined on $N$ lattice sites that only conserves fermion number parity.  In this case, the expansion of an $n$-point function in terms of two-point correlators will include contributions from both the regular $C_{nm} \equiv \braket{c^{\dagger}_{n}c_{m}}$ and `anomalous' correlation functions $F_{nm} \equiv \braket{c^{\dagger}_{n}c^{\dagger}_{m}}$. In order for the reduced density matrix to reproduce all observables in $A$, the entanglement Hamiltonian must also be a free-fermion Hamiltonian with pairing terms \cite{Peschel_SM}, i.e. of the form:
\begin{align}
\mathcal{H}_{E} = \sum_{n,m}\left[f_{nm}c^{\dagger}_{n}c_{m} + \Delta_{nm}c^{\dagger}_{n}c^{\dagger}_{m} + \mathrm{h.c.}\right]
\end{align}
As explained in \cite{Peschel_2_SM}, we may diagonalize the Hamiltonian by Bogoliubov transformation, so that the $\hat{C}$ and $\hat{F}$ matrices now satisfy the eigenvalue problem:
\begin{align}
(2\hat{C} - 2\hat{F} - 1)(2\hat{C} + 2\hat{F} - 1)\phi_{j} = \tanh^{2}\left(\frac{\epsilon_{j}}{2}\right)\phi_{j}
\end{align}
Equivalently, the entanglement Hamiltonian may be expressed in terms of Majorana operators $\gamma_{2n-1} \equiv c_{n} + c^{\dagger}_{n}$ and $\gamma_{2n} \equiv -i(c_{n} - c_{n}^{\dagger})$.  In this case, $\mathcal{H}_{E}$ will only contain terms bilinear in the Majorana operators, and may be written in the form:
\begin{align}
\mathcal{H}_{E} = \frac{i}{4}\sum_{n, m\in A}B_{nm}\gamma_{n}\gamma_{m}
\end{align}
with $\hat{B}$ a $2N \times 2N$ real, skew-symmetric matrix.  We may diagonalize the entanglement Hamiltonian by block-diagonalizing $\hat{B}$ by an appropriate orthogonal transformation and bringing it into canonical form:
\begin{align}
\mathcal{O}^{T}\hat{B}\mathcal{O} = \left(\begin{array}{cccccc}
0 & \epsilon_{1} &  &  &  & \\
-\epsilon_{1} & 0 &  &  &  & \\
 &  & 0 & \epsilon_{2} &  & \\
 &  & -\epsilon_{2} & 0 &  & \\
 & & & & & \ddots
\end{array}\right)
\end{align}
with $\mathcal{O}\in O(2N)$ so that the entanglement eigenvalues are $\pm\epsilon_{\ell}$. Taking the appropriate real linear combination of the Majorana operators $\{\gamma_{n}\}$ given by the eigenvectors of $\hat{B}$ that diagonalize the entanglement Hamiltonian, it is possible to compute the two-point correlation function $\Gamma_{nm} \equiv \braket{\gamma_{n}\gamma_{m}}$ $(n \ne m)$ in the original ground-state. Similar to the correlation function for a free-fermion system conserving fermion number, this yields the result \cite{Peschel_2_SM} that $\Gamma$ solves the eigenvalue problem:
\begin{align}
\hat{\Gamma}\phi_{\ell} = \pm\tanh\left(\frac{\epsilon_{\ell}}{2}\right)\phi_{\ell}
\end{align}
Therefore, in the limit $\epsilon_{\ell}\rightarrow 0$, we see that $2\hat{\Gamma}\phi_{\ell } \approx \pm\epsilon_{\ell}\phi_{\ell}$, so that the two-point function for Majorana fermions in the original ground-state is equivalent to the entanglement Hamiltonian when acting on the low-lying ($\epsilon_{\ell}\rightarrow 0$) states in the entanglement spectrum.

\section{The Two-Point Correlation Function for the Kitaev Majorana Chain}
We now compute the two-point function for Majorana fermions in the ground-state of the Kitaev Hamiltonian \cite{Kitaev_SM}
\begin{align}
H = &-w\sum_{n}\left(c^{\dagger}_{n+1}c_{n} + \text{h.c.}\right) + \sum_{n}\left(\Delta c^{\dagger}_{n+1}c^{\dagger}_{n} + \text{h.c.}\right)\nonumber\\ &- \mu\sum_{n}\left(c^{\dagger}_{n}c_{n} - \frac{1}{2}\right)
\end{align}
with $w = \Delta \in \mathbb{R}$. The complex fermion operators satisfy canonical anti-commutation relations $\{c_{n}, c_{m}\} = \{c_{n}^{\dagger}, c_{m}^{\dagger}\} = 0$ and  $\{c_{n}, c^{\dagger}_{m}\} = \delta_{nm}$.  In the topologically non-trivial phase, the ground-state of the Hamiltonian on an $N$-site chain with periodic boundary conditions takes the form of a BCS wavefunction with odd fermion parity:
\begin{align}
\ket{\Psi} = \prod_{k\ne 0}\left[u_{k} + v_{k}\cdot a^{\dagger}_{k}a^{\dagger}_{-k}\right]a^{\dagger}_{0}\ket{0}
\end{align}
where the parameter $u_{k}$ is given by:
\begin{align}
u_{k}^{2} = \frac{1}{2}\left(1 + \frac{2w\cos k + \mu}{E(k)}\right) 
\end{align}
with $u_{k}^{2} + v_{k}^{2} = 1$. Here, the quantity $E(k) = \sqrt{(2w\cos k + \mu)^{2} + 4w^{2}\sin^{2}k}$ is the bulk dispersion \cite{Kitaev_SM}.  We note that the non-vanishing correlation functions for the ground-state take the form:
\begin{align}
&\langle \Psi\,|\,c_{p}^{\dagger}c_{k}^{\dagger}\,|\,\Psi\rangle = \delta_{k,-p}\cdot u_{k}v_{k}\\
&\langle \Psi\,|\, c^{\dagger}_{k}c_{p}\,|\,\Psi\rangle = \delta_{k,p} \cdot v_{k}^{2}
\end{align}

We now explicitly compute the two-point correlation functions for the Majorana fermions in the ground-state of the clean Kitaev chain.  The real-space correlation functions for the complex fermions may be written by performing a Fourier transform:
\begin{align}
&\langle c^{\dagger}_{n} c^{\dagger}_{m}\rangle = \frac{1}{N}\sum_{k}e^{ik(m-n)}u_{k}v_{k}\\ &\langle c^{\dagger}_{n} c_{m}\rangle = \frac{1}{N}\sum_{k}e^{ik(m-n)}v_{k}^{2}
\end{align}
Now, we define the Majorana operators $\gamma_{2n}$ and $\gamma_{2n-1}$ for $n = 1,\ldots,N$ as $\gamma_{2n-1} \equiv c_{n} + c^{\dagger}_{n}$ and $\gamma_{2n} \equiv -i(c_{n} - c_{n}^{\dagger})$ so that the operators satisfy anti-commutation relations $\{\gamma_{n}, \gamma_{m}\} = 2\delta_{nm}$. When we perform a random partitioning of our system, we will be tracing over physical lattice sites. For the topologically non-trivial ground-state, this tracing procedure will produce dangling Majorana modes at the edges of connected clusters in the $A$ subsystem.  To determine the entanglement Hamiltonian obtained from a random partitioning, we are interested in computing the two-point function between Majorana fermions on odd and even Majorana sites that are separated by physical lattice sites belonging to the $B$ subsystem.  From the two-point functions for the complex fermions, we find that this correlation function for the Majorana fermions is given by the expression
\begin{align}\label{eq:Two_Pt_Function}
\langle \gamma_{2n}\gamma_{2m+1} \rangle = -i\delta_{n,m+1} + \frac{i}{N}\sum_{k} e^{ik(n-m-1)}\cdot v_{k}^{2}
\end{align}

We now define the quantity $\eta \equiv \mu/2w$.  In the limit $|\eta| << 1$, (near the Kitaev limit, as explained in the main text) we expand $v_{k}^{2}$ about $\eta = 0$, and extract the first non-trivial contribution to the two-point function $\braket{\gamma_{2n}\gamma_{2m+1}}$ to lowest order in $\eta$.  Note that:
\begin{align}
v_{k}^{2} = \frac{1}{2} - \frac{\eta + \cos k}{2\sqrt{1 + 2\eta\cos k + \eta^{2}}}
\end{align}
We now expand the above expression about $\eta = 0$. At each order in $\eta$, we wish to extract the term appearing in the expansion with the largest power of $\cos k$; this is the term that, when substituted into the sum appearing in (\ref{eq:Two_Pt_Function}) will provide the first non-zero contribution to the two-point function for a certain pair of well-separated Majorana fermions.  At $O(\eta^{\ell})$, we find that the term appearing in the expansion of $v_{k}^{2}$ with the highest power of $\cos k$ takes the form
\begin{align}
&\left( \frac{1}{2}\cdot\frac{3}{2}\cdots\frac{2\ell - 1}{2}\right)\frac{(-1)^{\ell + 1}(2\cos k + 2\eta)^{\ell}}{2(1 + 2\eta \cos k + \eta^{2})^{2\ell + 1}}\Bigg |^{\eta = 0} \frac{\eta^{\ell}}{\ell\,!} \cos k \nonumber\\
&= (-1)^{\ell+1}\frac{2^{\ell - 1}\eta^{\ell}}{\sqrt{\pi}}\frac{\Gamma\left(\ell + \frac{1}{2}\right)}{\Gamma(\ell + 1)}(\cos k )^{\ell + 1}
\end{align} 
where $\Gamma(x)$ is the gamma function. Here we have made use of the following identity:
\begin{align}
(2\ell - 1)!! = 1\cdot 3\cdot 5 \cdots (2\ell - 1) = \frac{2^{\ell}}{\sqrt{\pi}}\Gamma\left(\ell + \frac{1}{2}\right)
\end{align}
Now, substituting this term into the sum appearing in (\ref{eq:Two_Pt_Function}) for the two-point function for the Majorana fermions, we see that
\begin{widetext}
\begin{align}
&(-1)^{\ell+1}\frac{2^{\ell - 1}\eta^{\ell}}{\sqrt{\pi}}\frac{\Gamma\left(\ell + \frac{1}{2}\right)}{\Gamma(\ell + 1)}\frac{i}{N}\sum_{k}e^{ik(n-m-1)}(\cos k )^{\ell + 1} = (-1)^{\ell+1}\frac{\eta^{\ell}}{2\sqrt{\pi}}\frac{\Gamma\left(\ell + \frac{1}{2}\right)}{\Gamma(\ell + 1)}\frac{i}{N}\sum_{k}e^{ik(n-m-1)}\left[e^{ik(\ell + 1)} + \cdots\right]\nonumber\\
&= i (-1)^{\ell + 1} \frac{\eta^{\ell}}{2\sqrt{\pi}}\frac{\Gamma\left(\ell + \frac{1}{2}\right)}{\Gamma(\ell + 1)} \delta_{m-n, \ell} + \Big[\text{shorter-range couplings, i.e. terms proportional to $\delta_{m-n, s}$ with $s < \ell$}\Big]
\end{align}
\end{widetext}
Therefore, we observe that near the Kitaev limit, the above term provides the lowest-order contribution to the two-point function $C(\ell) \equiv \braket{\gamma_{2n}\gamma_{2n + 2\ell + 1}}$, with $\ell \ge 0$ i.e.
\begin{align}
C(\ell) = (-1)^{\ell + 1} \frac{\eta^{\ell}}{2\sqrt{\pi}}\frac{i\,\Gamma\left(\ell + \frac{1}{2}\right)}{\Gamma(\ell + 1)} + O(\eta^{\ell + 1})
\end{align}
This is precisely the two-point correlation function that determines the coupling in the entanglement Hamiltonian between two Majorana fermions at the edges of adjacent chains in the $A$ subsystem that are separated by $\ell$ physical lattice sites belonging to the $B$ subsystem. Our calculation shows that these dangling Majorana modes in the $A$ subsystem, when separated by $\ell$ physical lattice sites, will have an $O(\eta^{\ell})$ coupling.  We also see that when $\ell = 0$, we have $|2 C(\ell)| = 1$, in agreement with the dimensionless coupling between two nearest-neighbor Majorana fermions belonging to different physical lattice sites close to the Kitaev limit.  Finally, when $\ell$ is sufficiently large, $\Gamma(\ell + 1/2)/\Gamma(\ell + 1) \rightarrow 1/\sqrt{\ell}$, so that:
\begin{align}
 C(\ell) \sim \frac{\eta^{\ell}}{\sqrt{\ell}}
 \end{align}
at long distances.


\begin{thebibliography}{1}

\bibitem{levinwen}
M. Levin and X.G. Wen, Phys. Rev. Lett. {\bf 96}, 110405 (2006).

\bibitem{Kitaev_Preskill} A. Kitaev and J. Preskill, Phys. Rev. Lett. {\bf 96} 110404 (2006).

\bibitem{Li_Haldane} Li, H. and Haldane, F. D. M., Phys. Rev. Lett. {\bf 101}, 010504 (2008).

\bibitem {Edge_1} A. Chandran, M. Hermanns, N. Regnault, and B.A. Bernevig, Phys. Rev. B {\bf 84}, 205136 (2011).

\bibitem{Edge_2} X-L Qi, H. Katsura and A. W. W. Ludwig, Phys. Rev. Lett. {\bf 108}, 196402 (2012).

\bibitem{Edge_3} B. Swingle and T. Senthil, Phys. Rev. B {\bf 86}, 045117 (2012).

\bibitem{Edge_4} F. Pollmann, A. M. Turner, E. Berg and M. Oshikawa, Phys. Rev. B {\bf 81}, 064439 (2010).

\bibitem{Edge_5} L. Fidkowski, Phys. Rev. Lett. {\bf 104}, 130502 (2010).

\bibitem{Edge_7} J. Dubail and N. Read, Phys. Rev. Lett. {\bf 107}, 157001 (2011).


\bibitem{Hsieh_Fu} T. Hsieh and L. Fu, Phys. Rev. Lett. {\bf 113}, 106801 (2014).

\bibitem{Hsieh_Fu_Qi} T. Hsieh, L. Fu and Xiao-Liang Qi, Phys. Rev. B {\bf 90}, 085137 (2014).

\bibitem{Zhang} W. J. Rao, X. Wan and G. Zhang, Phys. Rev. B 90, 075151 (2014). 

\bibitem{santos} R. Santos, arXiv:1408.1716

\bibitem{hatsugai} T. Fukui and Y. Hatsugai, Journal of the Physical Society of Japan, {\bf 83}, 113705 (2014). 

\bibitem{Borchmann} J. Borchmann, A. Farrell, S. Matsuura, and T. Pereg-Barnea, arXiv:1407.5980v1 [cond-mat.str-el] (2014).

\bibitem{Huse_Motrunich_Damle} O. Motrunich, K. Damle, and D. A. Huse, Phys. Rev. B {\bf 63}, 224204 (2001).

\bibitem{Chalker_Coddington} J. T. Chalker and P. D. Coddington, J. Phys. C {\bf 21}, 2665 (1988).

\bibitem{Kitaev} A. Kitaev, arXiv:0010440v2 [cond-mat] (2000).


\bibitem{Peschel} I. Peschel, J. Phys. A {\bf 36}, L205 (2003).

\bibitem{Peschel_2} I. Peschel and V. Eisler, J. Phys. A {\bf 42}, 504003 (2009).


\bibitem{Fulga} I. C. Fulga, B. van Heck, J. M. Edge, and A. R. Akhmerov, Phys. Rev. B {\bf 89} 155424 (2014).

\bibitem{Fisher} D. S. Fisher, Phys. Rev. B {\bf 51}, 6411 (1995).

\bibitem{Yang} R. A. Hyman and K. Yang, Phys. Rev. Letters {\bf 78}, 1783 (1997).

\bibitem{Classification_1} A. P. Schnyder, S. Ryu, A. Furusaki, and A. W. W. Ludwig, Phys. Rev. B {\bf 78}, 195125 (2008).

\bibitem{Classification_2} A. Kitaev, arXiv:0901.2686 (2009).

\bibitem{Supp_Material} Supplementary Material
\end{thebibliography}

\begin{thebibliography}{1}

\bibitem{Peschel_SM} I. Peschel, J. Phys. A {\bf 36}, L205 (2003).
\bibitem{Peschel_2_SM} I. Peschel and V. Eisler, J. Phys. A {\bf 42}, 504003 (2009).
\bibitem{Kitaev_SM} A. Kitaev, arXiv:0010440v2 [cond-mat] (2000).

\end{thebibliography}
\end{document}